\documentclass{pasj00}
\draft

\newcommand{\kt}{\ensuremath{k_{\rm{B}}T}}
\newcommand{\lx}{\ensuremath{L_{\rm{X}}}}

\newcommand{\fx}{\ensuremath{F_{\rm{X}}}}
\newcommand{\nh}{\ensuremath{N_{\rm H}}}

\SetRunningHead{Y.~Hyodo et al.}{X-ray Spectrum of a Peculiar Star} 
\Received{2007/6/12}
\Accepted{2007/8/31}
\begin{document}
\title{Suzaku X-Ray Spectroscopy of a Peculiar Hot Star \\in the 
  Galactic Center Region}
\author{Yoshiaki \textsc{Hyodo},\altaffilmark{1} 
Masahiro \textsc{Tsujimoto},\altaffilmark{2,3}\thanks{Chandra Fellow} 
Katsuji \textsc{Koyama},\altaffilmark{1} 
Shogo \textsc{Nishiyama},\altaffilmark{4}\\ 
Tetsuya \textsc{Nagata},\altaffilmark{5} 
Itsuki \textsc{Sakon},\altaffilmark{6} 
Hiroshi \textsc{Murakami},\altaffilmark{7} 
and 
Hironori \textsc{Matsumoto}\altaffilmark{1} 
}
\altaffiltext{1}{Department of Physics, Graduate School of Science, 
Kyoto University, \\Kita-shirakawa Oiwake-cho, Sakyo, Kyoto 606-8502}
\altaffiltext{2}{Department of Astronomy \& Astrophysics, Pennsylvania 
State University \\
525 Davey Laboratory, University Park, PA 16802, USA}
\altaffiltext{3}{Department of Physics, Rikkyo University, 3-34-1, 
Nishi-Ikebukuro, Toshima, Tokyo 171-8501}
\altaffiltext{4}{National Astronomical Observatory of Japan, 2-21-1, Osawa, 
Mitaka, Tokyo 181-8588}
\altaffiltext{5}{Department of Astronomy, Graduate School of Science, 
Kyoto University, \\Kita-shirakawa Oiwake-cho, Sakyo, Kyoto 606-8502}
\altaffiltext{6}{Department of Astronomy, School of Science, 
the University of Tokyo, 7-3-1, Hongo, Bunkyo, Tokyo 113-0033}
\altaffiltext{7}{Institute of Space and Astronautical Science, 3-1-1, 
Yoshinodai, Sagamihara, Kanagawa, 229-8510}
\email{hyodo@cr.scphys.kyoto-u.ac.jp}
\maketitle

\begin{abstract}
  We present the results of a Suzaku study of a bright point-like source 
  in the 6.7~keV intensity map of the Galactic center region. 
  We detected an intense Fe\emissiontype{XXV} 6.7~keV line with an 
  equivalent width of $\sim$\,1~keV as well as emission lines of 
  highly ionized Ar and Ca from a spectrum obtained by the X-ray 
  Imaging Spectrometer. 
  The overall spectrum is described very well by a heavily absorbed 
  ($\sim$\,2$\times$10$^{23}$~cm$^{-2}$) thin thermal plasma model 
  with a temperature of 3.8\,$\pm$\,0.6~keV and a luminosity 
  of $\sim$\,3$\times$10$^{34}$~erg~s$^{-1}$ (2.0--8.0~keV) at 8~kpc. 
  The absorption, temperature, luminosity, and the 6.7~keV line 
  intensity were confirmed with the archived XMM-Newton data. 
  The source has a very red ($J-K_{\rm s}=8.2$~mag) infrared 
  spectral energy distribution (SED), which was fitted by 
  a blackbody emission of $\sim$\,1000~K attenuated by a visual 
  extinction of $\sim$\,31~mag. 
  The high plasma temperature and the large X-ray luminosity
  are consistent with a wind-wind colliding Wolf-Rayet binary. 
  The similarity of the SED to those of the eponymous Quintuplet 
  cluster members suggests that the source is a WC-type source.  
\end{abstract}
\section{Introduction}
The distribution of the 6.7~keV X-ray emission line along the Galactic
plane is strongly peaked at the Galactic center \citep{koyama89}. 
The line originates from a K$\alpha$ transition of highly
ionized iron. Together with its large penetrating power delving
through a $\approx$\,10$^{24}$~H~cm$^{-2}$ extinction, the emission is an
indispensable tool to trace the high-energy activity and to reveal the
X-ray source demographics in a heavily attenuated environment, such as
the Galactic center region, where the enormous
density of gas and dust, the strong magnetic fields, and the
existence of a super massive black hole influence every aspect of
the transmigration of energy and matter \citep{morris96}.

Both the extended emission and the ensemble of numerous discrete
sources contribute to the 6.7~keV emission. 
The nature of the extended emission was identified as a hot plasma 
having a temperature of $\sim$\,6.5~keV, distributed across $\sim$\,250~pc
(\cite{koyama89}, \yearcite{koyama07c}). 
For discrete sources, several different X-ray populations with 
thermal emission play dominant roles.

Cataclysmic variables (CV), binaries of a white dwarf and 
predominantly a late-type dwarf star, 
are major contributors to the 6.7~keV line, considering its 
large population and thermal plasma with emission lines 
(\cite{ezuka99}; \cite{muno03}, \yearcite{muno04b}). 
Low-mass young stellar objects (YSO) or pre--main-sequence late-type 
sources are other types of candidates. 
They have elevated X-ray activity compared to their main-sequence 
phases, and show a hard
thermal X-ray emission with the 6.7~keV line both during the occasional flare 
and quiescent phases \citep{koyama96a,tsuboi98,imanishi01,ozawa05}. 
They are expected to be numerous in the Galactic center region, 
where $\sim$\,10\% of the star formation of the entire Galaxy takes 
place \citep{figer04}. 

In addition to these late-type populations, early-type sources are
also expected to contribute to the 6.7~keV emission. 
Main-sequence field O-type stars do not show hard X-ray emission 
with their plasma temperatures below $\sim$\,1~keV \citep{berghoefer97}, 
but recent observations reveal that some main-sequence 
O-type stars in OB associations show hard thermal X-ray emissions of 
$\gtsim$\,2~keV (\cite{tsujimoto07b} and references therein). 
Intense 6.7~keV line emissions are confirmed from such sources 
(\cite{albacete03}, \yearcite{albacete07}; \cite{broos07}; \cite{hyodo08}). 
Early-type sources are extremely rare compared to 
late-type sources in number, but their overwhelmingly large luminosity 
can be comparable to the integrated luminosity of numerous late-type sources. 
In the Orion Nebula Cluster,
for instance, an O6V star ($\theta^{1}$ Ori\,C) alone accounts for
$\sim$\,34\% of the total hard-band X-ray emission integrated over the
entire Chandra field of view containing $>$\,1000 YSOs \citep{feigelson05}

Early-type stars become even brighter in X-rays when they evolve 
off the main-sequence track. 
Wolf-Rayet (WR) stars and Luminous Blue Variables
(LBVs) comprise the brightest X-ray population of thermal point-like
sources with a typical hard-band luminosity of 
$\sim$\,10$^{34}$~erg~s$^{-1}$ and a plasma temperature of $\gtrsim$\,2~keV 
\citep{koyama94,tsuboi97,portegies02b,wang06}. 
In fact, the brightest stellar X-ray source not powered  
by accretion reported to date in our Galaxy 
is a WR star of a spectral type of WN7--8 in the Arches cluster 
(A1\,S in \cite{yusef02, law04, wang06,tsujimoto07a}). 
The source has the 0.2--10.0~keV luminosity exceeding 
$\sim$\,10$^{35}$~erg~s$^{-1}$ with a conspicuous 6.7~keV 
emission.
Moreover, the X-ray spectra of these stars commonly show strong metallic 
emission lines against continuum including the 6.7~keV line, which is
a consequence of extreme hydrogen depletion caused by the mass loss
and hydrogen burning \citep{hucht1986}.

\smallskip

Although early-type stars known in the Galactic center region are
concentrated to the three massive young star clusters --- 
the Arches \citep{nagata95, cotera96, serabyn98, figer99b}, 
the Quintuplet \citep{kobayashi83, okuda90, nagata90, figer99a}, 
and the Central cluster \citep{becklin68, krabbe95, ghez05} ---,  
it is quite natural to expect that a much larger number of early-type 
stars remain unidentified. 
\citet{portegies01} claimed that the number of
young massive star clusters may exceed 50 in the Galactic center
region. The inconsistency with the observed value indicates that these
clusters are too obscured to be visible in the optical and infrared bands, 
or that the cluster members are dissipated before 
the earliest member reaches its end, or that that number is a 
gross overestimate. 
\citet{kim99} showed that the strong tidal disruption in 
the Galactic center region makes massive star clusters become unbound on a
very short time scale comparable to the lifetime of an O star. 
Isolated early-type stars may be distributed throughout the region, unlike
the other parts of the Galaxy where they are found in associations.

\begin{table*}[t]
  \begin{center}
    \caption{Observation log.}
    \label{tabl:obslog}
    \begin{tabular}{ccccccr}
      \hline
      Start Date&Observatory&ObsID&R.\,A. & Decl.&$t_{\rm exp}$\footnotemark[$*$]&
      $\theta$\footnotemark[$\dagger$]\\
      &&&\multicolumn{2}{c}{(J2000.0)}&(ks)&(\arcmin)\\
      \hline
      2000-09-23&XMM-Newton&0112970201&\timeform{17h47m21s}&--\timeform{28D09'02''}&11&11.7\\
      2000-10-27&      Chandra&   1036      &\timeform{17h47m22s}&--\timeform{28D11'36''}&35&10.5\\
      2001-07-16&      Chandra&   2271      &\timeform{17h47m28s}&--\timeform{28D16'29''}&10&9.5\\
      2001-07-16&      Chandra&   2274      &\timeform{17h46m42s}&--\timeform{28D10'23''}&10&5.7\\
      2001-07-16&      Chandra&   2285      &\timeform{17h46m18s}&--\timeform{28D20'23''}&10&7.9\\
      2003-03-12&XMM-Newton&0144220101&\timeform{17h47m23s}&--\timeform{28D09'15''}&34&10.4\\
      2006-09-21&      Suzaku&501040010     &\timeform{17h46m46s}&--\timeform{28D22'51''}&70&7.4\\
      2006-09-24&      Suzaku&501040020     &\timeform{17h46m46s}&--\timeform{28D22'51''}&50&7.4\\
      \hline
      \multicolumn{4}{@{}l@{}}{\hbox to 0pt{\parbox{135mm}{\footnotesize
            \par\noindent
	    \footnotemark[$*$] Exposure time after removing periods with 
	    high background level. 
	    The exposure times for the XMM-Newton observations refer to 
	    those obtained with the EPIC-pn camera. 
	    \par\noindent
	    \footnotemark[$\dagger$] Angular distance of the source 
	    from the optical axis.
	  }\hss}}
    \end{tabular}
  \end{center}
\end{table*}

Some early attempts have been made to discover unidentified early-type
stars in the Galactic center region with the combination of X-ray and 
infrared (IR) observations. 
\citet{muno06a} made trailblazing observations of radio, IR, and X-ray 
to reveal the population of young massive stars in the Galactic center region. 
They detected strong Br$\gamma$ and He\emissiontype{I} lines from two 
sources, and classified them as either Of or candidate LBV stars. 
\citet{mauerhan07} also conducted
\textit{K}-band spectroscopy of six bright hard X-ray sources with
very red IR colors. 
Two of them, with IR colors of
\textit{J}--\textit{K}$=4$--5~mag, show broad hydrogen and helium
emission lines characteristic of evolved O-type stars. 
They are classified as a Wolf-Rayet star of a spectral type of WN6 
and an O-type Ia supergiant. 
Likewise, \citet{mikles06} performed infrared spectroscopy of 
CXO\,J174536.1--285638, which is one of new Chandra sources discovered 
by \citet{muno03}. 
They detected P Cygni profiles of He\emissiontype{II} lines of 
a 170~km~s$^{-1}$ wind. 
The spectral features in the X-ray and IR bands are most consistent 
with the colliding wind binary system $\eta$ Car. 
This source is of particular interest in terms of a strong 
Fe\emissiontype{XXV} emission line, 
which is also prominent in our source, presented below. 
We see a high prospect of these methods to find similar sources en masse. 

Near-IR spectroscopy may confront a challenge for even redder sources. 
In addition to a $\sim$\,30~mag visual extinction ubiquitous
toward the Galactic center region \citep{catchpole90,schultheis99}, 
evolved early-type stars are behind
an additional local extinction by their own mass loss. 
Photospheric emission is reprocessed into longer wavelength 
radiation in the optically thick circumstellar matter. 
The IR spectra turn out to be featureless by strong 
veiling \citep{figer99a,crowther06}. 
For such sources, the spectroscopy of the hard X-ray emission directly 
from the vicinity of the star is the only practical tool. 
These stars contribute to the 6.7~keV line emission,
and the diagnosis of the line emission helps to identify their nature.

\medskip

Here, we present the results of a Suzaku \citep{mitsuda07} study of an X-ray 
point-like source located at a $\sim$\,100~pc projected distance from 
the Galactic center. 
The source is exceptionally bright in the 6.7~keV map of the region, 
and is redder (\textit{J}--\textit{K}\,$>$\,8~mag) than those 
studied in previous studies. Aided by the large effective area of Suzaku's 
telescope to produce spectra having high photon statistics, 
we conducted a detailed hard-band spectroscopy of this source 
to identify its nature. 

\section{Observation and Data Reduction}

As a part of the Suzaku mapping campaign of the Galactic center region, 
a field containing the Sgr\,B2 molecular cloud was observed twice on 2006 
September 21--23 and 24--25 with an aim point at 
(RA, Dec)\,$=$\,(\timeform{17h46m46s}, --\timeform{28D22'51''}).
The observation log is given in table~\ref{tabl:obslog}. 
Suzaku observations produce two simultaneous data sets using 
the X-ray Imaging Spectrometer (XIS: \cite{koyama07a}) and 
the Hard X-ray Detector (HXD: \cite{kokubun07,takahashi07}). 
We concentrate on the former in this paper. 

The XIS is equipped with four X-ray charge coupled devices (CCDs) 
at the foci of four X--Ray Telescopes (XRT: \cite{serlemitsos07}). 
Each CCD chip has a format of 1024$\times$1024 pixels. 
The four XRTs are aligned to observe the same field of 
$\sim18\arcmin\times$18\arcmin\ with a half-power diameter (HPD) of 
1\farcm9--2\farcm3 and a pointing accuracy of up to $\sim$\,50\arcsec.
Three of the four chips (XIS0, 2, and 3) are front-illuminated (FI) CCDs 
superior in the hard band response covering the 0.4--12~keV energy range. 
The remaining one is a back-illuminated (BI) CCD superior in the soft band 
covering 0.2--12~keV. 
The total effective area is $\sim$\,550~cm$^2$ at 8~keV.

The energy resolution, initially of $\sim$130 eV at 6~keV in the full width 
at half maximum, is subject to degradation due to cosmic-ray 
radiation in the orbit. 
The XIS employs two techniques to calibrate and rejuvenate its spectral 
capability. 
One is standard radioactive sources. 
Two \atom{Fe}{}{55} sources are installed to illuminate two corners of 
each CCD to calibrate the absolute energy gain. 
The other is the spaced-row charge injection (SCI). 
Electrons are injected at one of every 54 rows, and are 
transferred through columns to sacrificially fill in traps caused 
by radiation. 
This reduces the number of trapped charges by X-ray events throughout 
the transfer, thus improving the energy resolution \citep{bautz04, koyama07a}.

The observations were conducted using the normal clocking mode 
with a frame time of 8~s. The SCI technique was used. 
Data process version 1.2\footnote{See 
http://www.astro.isas.jaxa.jp/suzaku/process/history/v1225.html.} 
was retrieved and events were 
removed during the South Atlantic Anomaly passages and at 
earth elevation angles below 3~degrees. 
After the filtering, the combined net integration time is $\sim$\,120~ks.

Because the calibration database for SCI observations is not yet released 
as of this writing, we used data obtained during the ground calibration 
with a charge-transfer efficiency of 1. 
Using the Mn K$\alpha$ line from the calibration sources 
$\left(\atom{Fe}{}{55}\right)$, 
we measured the systematic gain offset as the deviation from 
the theoretical value at 5.8951 keV \citep{bearden67, krause79}. 
The offsets were found to be within $\sim$\,10~eV for the FI 
and $\sim$\,30~eV for the BI chips.

\begin{table*}[!htb]
  \begin{center}
    \caption{Best-fit spectral parameters.\footnotemark[$*$]}
    \label{tabl:apec}
    \begin{tabular}{cccc}
      \hline
      Parameter&Unit&Suzaku&XMM-Newton\\
      \hline
      \nh&($10^{23}$~cm$^{-2}$)&2.4$^{+0.3}_{-0.2}$&2.3$^{+0.4}_{-0.3}$\\
      \kt&(keV)&3.8$^{+0.5}_{-0.6}$&3.8$^{+0.7}_{-0.7}$\\
      $Z_{\rm Ar}$& (solar)&8.7$^{+6.3}_{-4.9}$&2.9$^{+7.1}_{-2.9}$\\
      $Z_{\rm Ca}$& (solar)&3.2$^{+2.3}_{-2.0}$&0.9$^{+2.8}_{-0.9}$\\
      $Z_{\rm Fe}$& (solar)&0.8$^{+0.1}_{-0.1}$&0.9$^{+0.2}_{-0.2}$\\
      \fx\footnotemark[$\dagger$]&($10^{-12}$~erg~s$^{-1}$~cm$^{-2}$)&
      1.00$^{+0.03}_{-0.03}$&1.05$^{+0.07}_{-0.07}$\\
      \lx\footnotemark[$\ddagger$]&($10^{34}$~erg~s$^{-1}$)&2.8&2.9\\      
      $\chi^2$/d.o.f.&&89.4/106&69.9/74\\
      \hline
      \multicolumn{2}{@{}l@{}}{\hbox to 0pt{\parbox{100mm}{\footnotesize
	    \par\noindent
	    \footnotemark[$*$] The uncertainty indicates the 90\% confidence
	    ranges of the fit. 
            \par\noindent
	    \footnotemark[$\dagger$] Energy flux in the 2.0--8.0~keV band.
	    \par\noindent 
	    \footnotemark[$\ddagger$] Absorption-corrected luminosity 
	    in the 2.0--8.0~keV band. 
	    A distance of 8~kpc is assumed. 
	  }\hss}} 
    \end{tabular} 
  \end{center} 
\end{table*} 

\section{Analysis}
\subsection{Images}
\begin{figure*}[!htb]
  \begin{center}
    \FigureFile(85mm,85mm){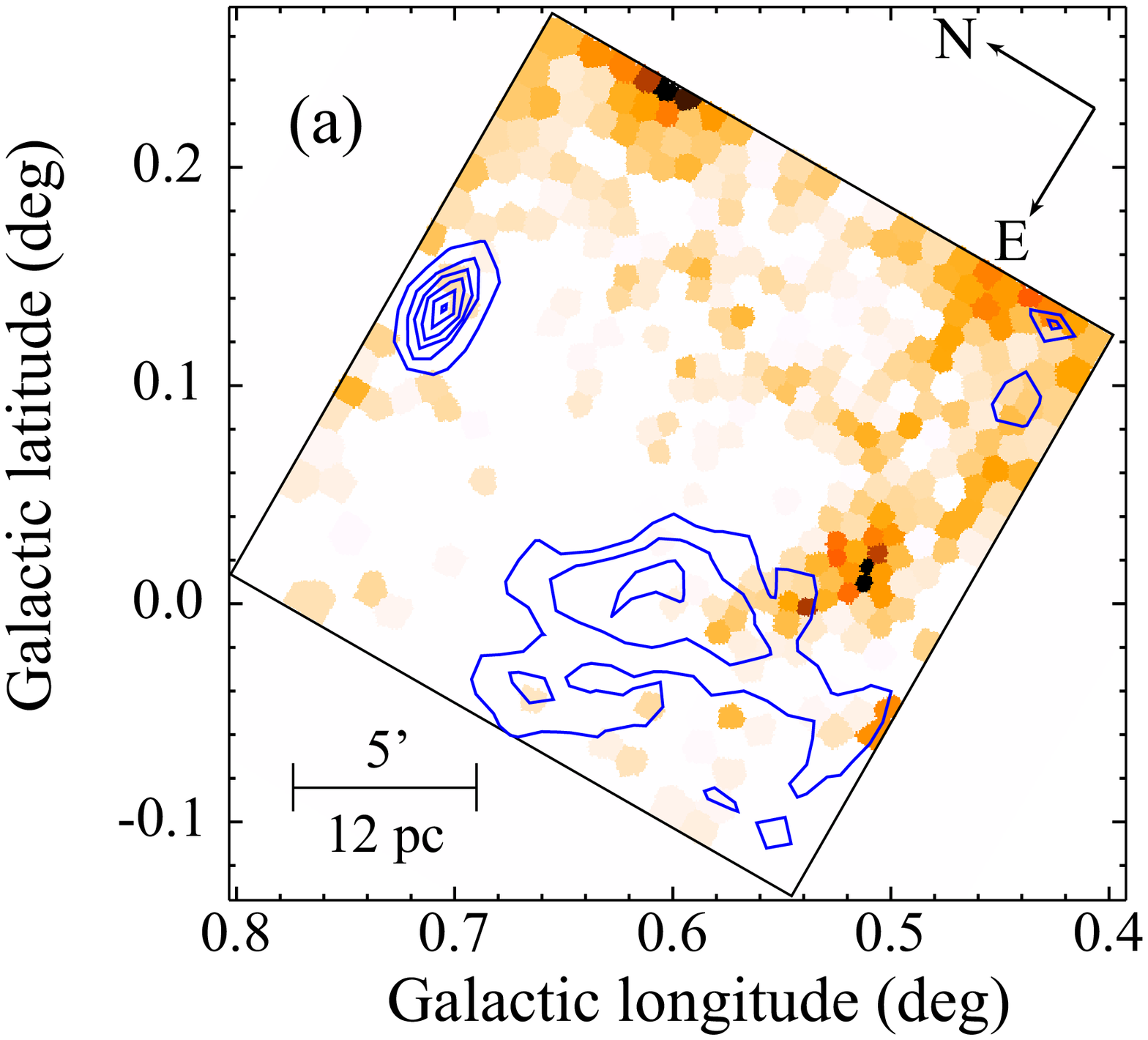}
    \FigureFile(85mm,85mm){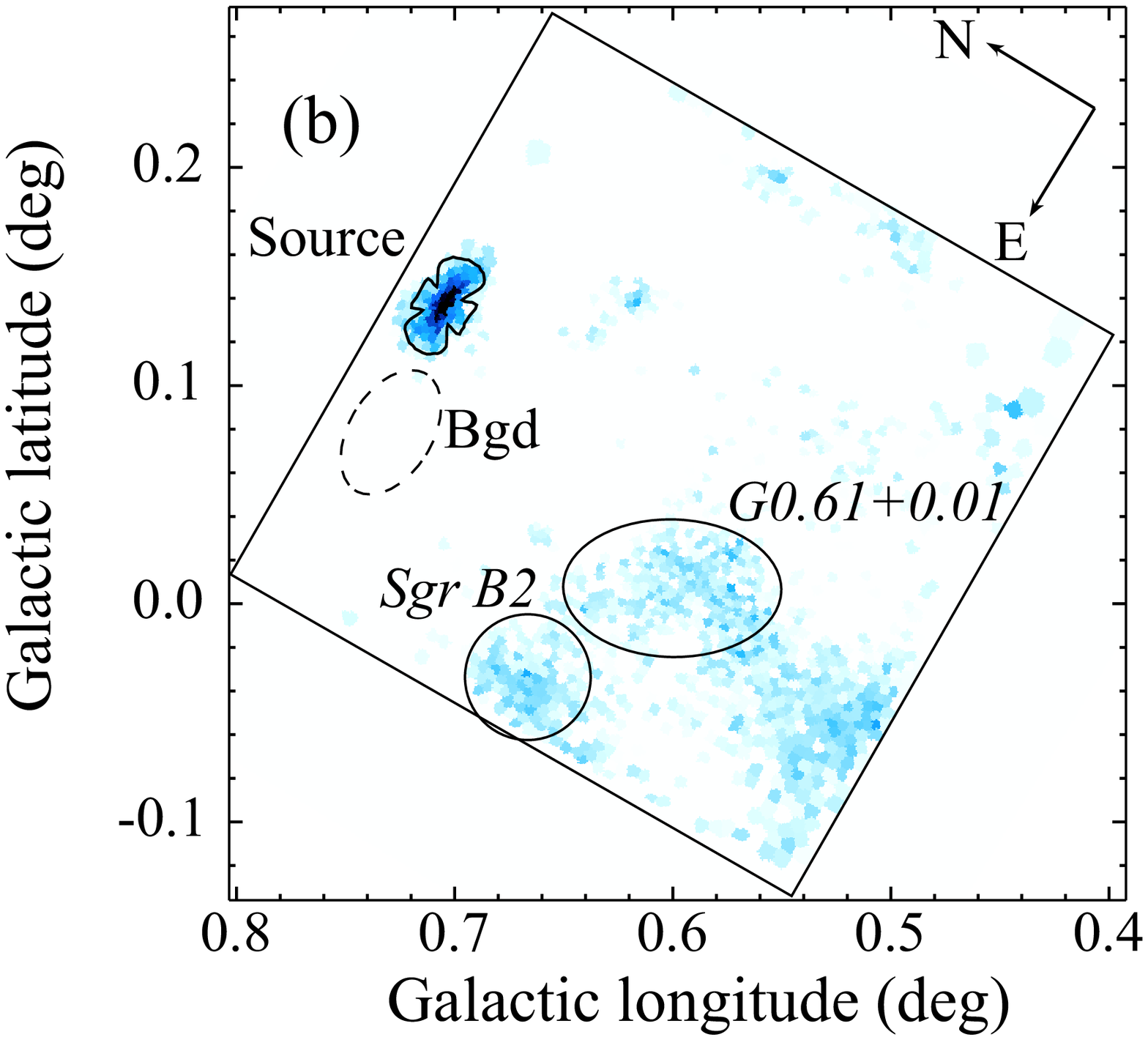}
  \end{center}
  \caption{XIS images in the (a) 0.7--2.0 and (b) 2.0--7.0~keV bands.
    The narrow-band intensity of 6.57--6.77~keV is shown in (a) 
    by contours. 
    The source and background regions are respectively shown by 
    a polygon and a dashed ellipse in (b). 
    Both images were processed (1) to mask calibration sources, 
    (2) to subtract non--X-ray background signals constructed from 
    night earth observations \citep{tawa08}, 
    (3) to correct for the vignetting and the non-uniformity caused 
    by the contamination material on the optical blocking filter,  
    (4) to resample adaptively to achieve a signal-to-noise ratio of 
    $>$\,8 at each bin using the weighted Voronoi tessellation algorithm
    (\cite{diehl06}, \cite{cappellari03}), and (5) to register the 
    coordinate using the Chandra counterpart of the point-like source.
  }\label{fg:image}
\end{figure*}
Figure \ref{fg:image} shows XIS images of the study field 
in the (a) 0.7--2.0 and (b) 2.0--7.0~keV bands. 
The overlaid contours in (a) are the narrow-band intensity in 6.75--6.77~keV. 
All four XIS images at the two observations are combined. 
Several distinctive features can be found in the hard band. 
Two extended sources are 
Sgr\,B2 (\cite{koyama96b}, \cite{murakami00}, \yearcite{murakami01}; 
\cite{koyama07b}) 
and G0.61--0.01 \citep{koyama07b}.
The brightest source, which is the main topic of this paper, 
is found close to the northern edge of the hard-band image. 
The source is point-like and an exceptionally intense 6.7~keV emitter 
in the image, and also in the vicinity found in the Suzaku wide-field map.

\subsection{Spectra}
We extracted the source and background events from the polygonal and 
elliptical regions, respectively (figure~\ref{fg:image}b). 
The source was superimposed upon the intense diffuse emission ubiquitous in 
the Galactic center region \citep{koyama96b}. 
In order to maximize the signal-to-noise ratio against the 
underlying emission, 
we simulated point spread functions at the source position 
with different enclosed energy fractions using a ray-tracing simulator 
(\texttt{xissim}: \cite{ishisaki07}). 
Consequently, a 60\% encircled energy polygon was chosen to 
accumulate source photons. 
The background region was selected from an ellipse adjacent to and 
at a similar off-axis angle with the source region.

\begin{figure}[!htb]
  \begin{center}
    \FigureFile(85mm,85mm){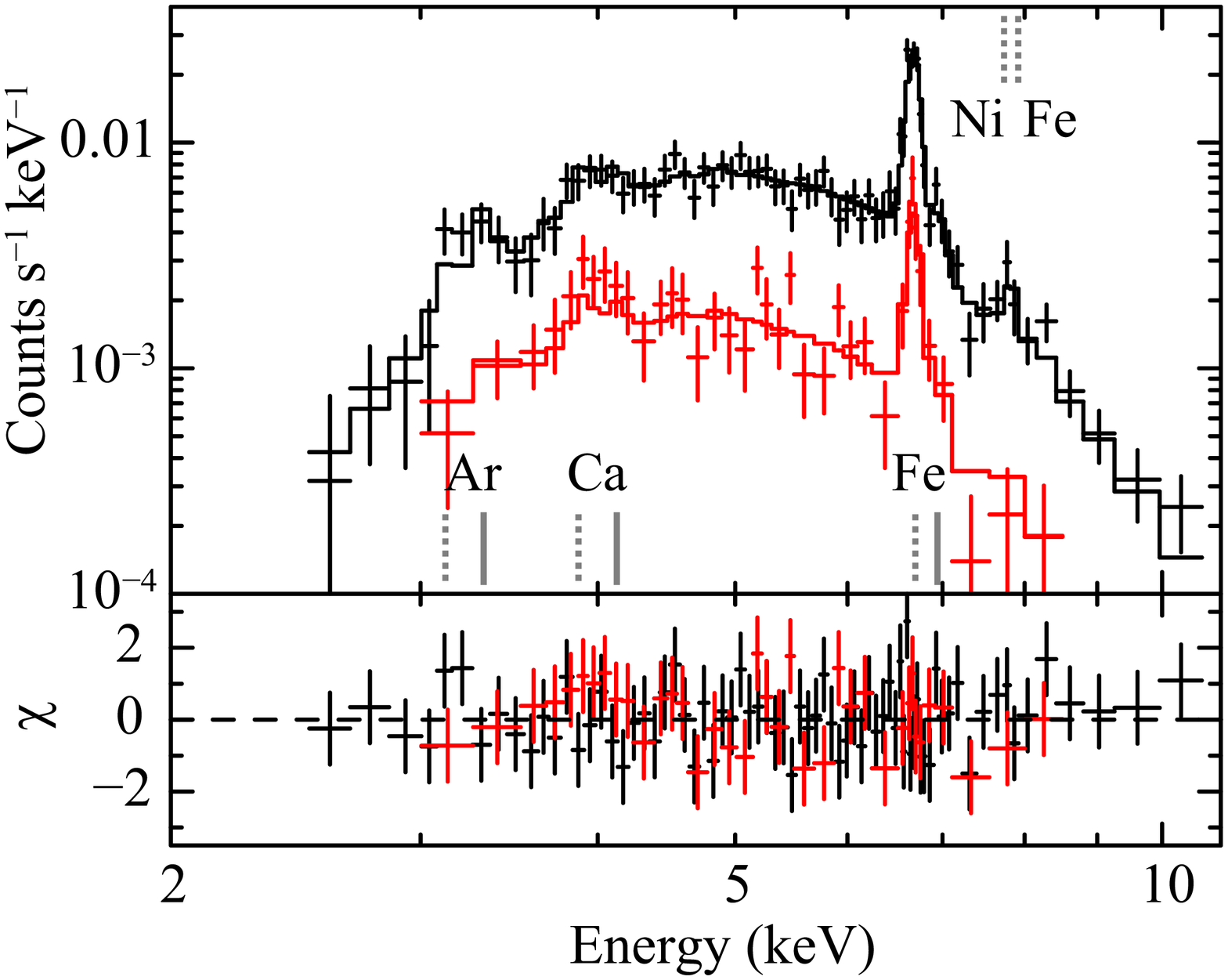}
  \end{center}
  \caption{Background-subtracted XIS spectra. BI spectrum is
    shown in red, while the merged FI spectrum is in black. The
    upper panel shows the data with crosses and the best-fit APEC
    models with solid lines, while the lower panel shows the residuals
    to the best-fit. Conspicuous emission lines of He-like ions 
    (gray dash lines) and H-like ions (gray solid lines) are labeled.
    }
  \label{fg:bi3fispec}
\end{figure}

Figure~\ref{fg:bi3fispec} shows the background-subtracted spectrum. 
Here, we generated the telescope response files using 
a ray-tracing simulator (\texttt{xissimarfgen}: \cite{ishisaki07}). 
We used the detector response files as of 2005 August, which best describe 
the spectral profiles of the calibration sources. 

The spectrum is characterized by several emission lines as well as a hard 
continuum extending to $\sim$\,10~keV with a strong soft-band cut-off. 
The lines are identified as K$\alpha$ emission from highly 
ionized (He-like and H-like) ions of Ar, Ca, and Fe. 
The spectrum was first fitted by a bremsstrahlung model with 
Gaussian lines attenuated by interstellar extinction \citep{morrison83}. 
The strongest line is He-like Fe K$\alpha$ at $6.66\pm0.01$~keV 
with an energy flux of $(2.5\pm0.3)\times10^{-13}$~erg~s$^{-1}$~cm$^{-2}$, 
which corresponds to an equivalent width (EW) of $950\pm100$~eV. 

In contrast, the line feature at 6.4~keV, if present, is very weak. 
We added a narrow Gaussian line at the energy and derived
a 90\% upper limit of 
1.5$\times10^{-14}$~erg~s$^{-1}$~cm$^{-2}$, which corresponds to an 
EW of 50~eV. 
The temperature derived from the continuum and those from the 
intensity ratios of He-like and H-like K$\alpha$ lines are consistent 
with $\sim$\,4~keV, indicating that the emission is thermal plasma 
in collisional equilibrium.

We then fitted the spectrum with an optically-thin thermal plasma 
model (APEC: \cite{smith01}) with interstellar 
extinction \citep{morrison83}. 
The abundances of Ar, Ca, and Fe relative to solar were free parameters, 
while the other elements were fixed to 1~solar abundance \citep{anders89}. 
A single-temperature model yielded an acceptable fit with 3.8~keV 
in plasma temperature ($k_{\rm{B}}T$), 
$1.0\times10^{-12}$~erg~s$^{-1}$~cm$^{-2}$ in X-ray flux ($F_{{\rm X}}$) 
in the 2.0--8.0~keV band 
and 2.4$\times10^{23}$~cm$^{-2}$ in hydrogen column extinction 
($N_{\rm{H}}$). 
The best-fit model and values are shown in figure~\ref{fg:bi3fispec} 
and table~\ref{tabl:apec}, respectively. 
No systematic deviation was found in the single temperature fit, 
requiring no extra components. 

To examine how the results are influenced by our choice of the 
background region, 
we repeated the same procedures with several different regions. 
All the best-fit parameters are consistent with each other, except for the 
X-ray flux, which suffers $\lesssim$\,10\% systematic uncertainty. 

\subsection{Chandra and XMM-Newton Counterparts} 
We retrieved the archived Chandra \citep{weisskopf02} and 
XMM-Newton \citep{jansen01} data to study the long-term behaviors of 
the source, and to locate its position more precisely using telescopes 
of smaller HPDs than that of Suzaku. 
Four Chandra observations using the Advanced CCD Imaging Spectrometer 
(ACIS: \cite{garmire03}) and two XMM-Newton observations using the 
European Photon Imaging Camera (EPIC: \cite{turner01, strueder01}), which is 
comprised of two MOS and a PN, were found to cover 
the Suzaku source (table~\ref{tabl:obslog}). 

Within the Suzaku positional uncertainty of 50\arcsec, 
we found only one Chandra source (CXOGC\,J174645.3--281546) in \citet{muno06b} 
and one XMM-Newton source (2XMMp\,J174645.2--281547) in the 
Second XMM-Newton Serendipitous Source 
Pre-release Catalogue, XMM-Newton Survey Science Centre (2006).
Also, hereafter we refer the Suzaku source as CXOGC\,J174645.3--281546. 

\begin{figure}[!htb]
  \begin{center}
    \FigureFile(85mm,85mm){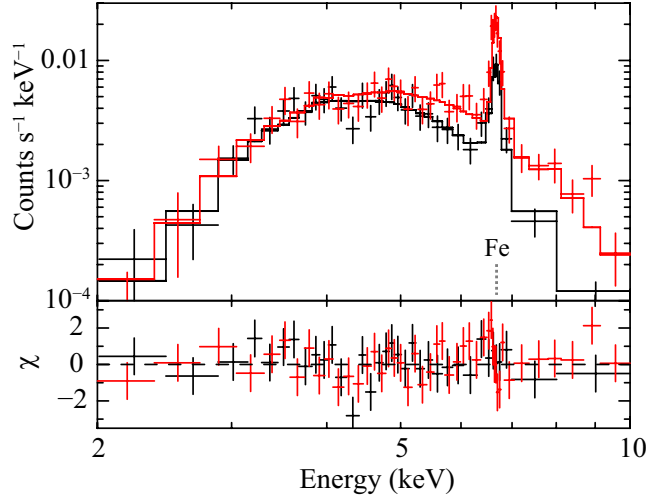}
  \end{center}
  \caption{Background-subtracted EPIC spectra. 
    The pn spectrum is shown in red, 
    while the merged MOS spectrum is in black.
    The symbols follow figure~\ref{fg:bi3fispec}.}
  \label{fg:epicspec}
\end{figure}

Since no spectroscopic and temporal behaviors are presented for both of the 
Chandra and XMM-Newton sources in the literature, 
we analyzed the data and present the results here. 
One of the two XMM-Newton observations showed a net exposure of $\sim$\,34~ks 
after removing data during high background. 
Spectra with high statistics were obtained from MOS and PN, for which 
we conducted spectral fits in a similar manner as with the Suzaku spectrum. 
The best-fit model and parameters are shown in figure~\ref{fg:epicspec} 
and table~\ref{tabl:apec}, respectively. 
Although the emission lines are less conspicuous in the XMM-Newton data, 
the Suzaku and XMM-Newton results are consistent with each other. 

For the remaining one XMM-Newton and four Chandra observations, 
the photon statistics were too poor for a detailed spectral analysis, 
due to short exposures, chip gaps, and 
large off-axis angles of the source position. 
We therefore applied the best-fit Suzaku model to derive their flux.

\begin{figure}[h]
  \begin{center}
    \FigureFile(80mm,80mm){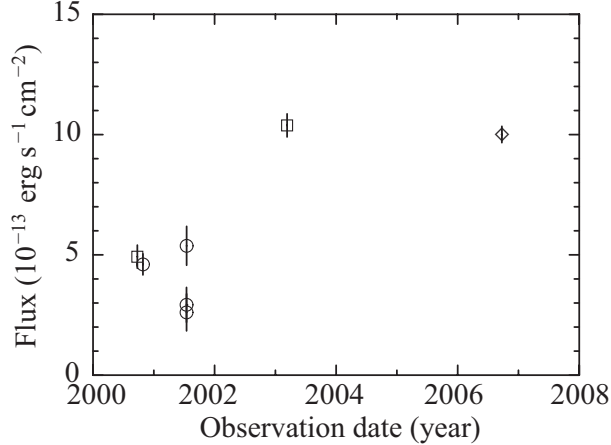}
  \end{center}
  \caption{Long-term trend of X-ray flux in the 2.0--8.0~keV using 
    XMM-Newton (squares), Chandra (circles), and Suzaku (a diamond).
    Error bars on the data points are plotted at the 90\% confidence level. 
  }
  \label{fg:long}
\end{figure}

Figure~\ref{fg:long} shows a long-term flux variation, 
which spans $\sim$\,6~years at seven different epochs. 
Although no flux variation within each observation is found, 
a long-term variation having a factor $\sim2$ is clearly found. 

\subsection{Longer Wavelength Data}
We retrieved other databases to characterize the multi-wavelength features 
of CXOGC\,J174645.3--281546. 
Because of the extreme extinction, the source is inaccessible in the 
optical bands. 
In the near-infrared (NIR) bands, we examined 
the Two Micron All-Sky Survey (2MASS: \cite{cutri03}; \cite{skrutskie06}) and 
the NIR Galactic center survey \citep{nishiyama06} using 
Simultaneous Infrared Imager for Unbiased Survey
(SIRIUS: \cite{nagashima99, nagayama03}) on the Infrared Survey Facility 
(IRSF) telescope.
In the mid-infrared (MIR) bands, we  
used the Midcourse Space Experiment (MSX: \cite{egan03}) and the Galactic 
Legacy Infrared Mid-Plane Survey Extraordinaire program (GLIMPSE: 
\cite{benjamin03}) using the Spitzer Space Telescope \citep{werner04}. 

\begin{figure}[h]
  \begin{center}
    \includegraphics[width=8cm,height=5.55cm]{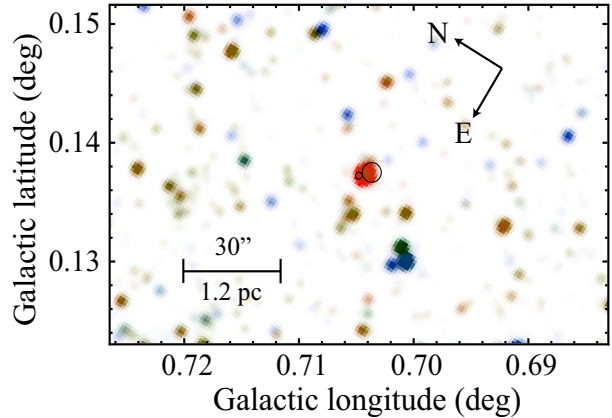}
  \end{center}
  \caption{NIR pseudo-color image using 2MASS $J$- (blue), $H$- (green), 
    and $K_{\rm s}$- (red) band images. 
    The source positions with the uncertainty 
    of Chandra and MSX are shown with a small and large circle, respectively.
  }
  \label{fg:irimage}
\end{figure}

Figure~\ref{fg:irimage} shows an NIR pseudo-color image, 
in which an isolated point-like source (2MASS J17464524--2815476 and 
MSX C6 G000.7036$+$00.1375) is found in the 2MASS and MSX images 
within the positional uncertainty range of the Chandra source. 
The source is very bright and red with 2MASS magnitudes of 
(\textit{J}, \textit{H}, \textit{K}$_{\rm{s}}$)\,$=$\,($>$\,15.4, 
10.0, $>$\,7.2) mag. 
The \textit{J}- and \textit{K}$_{\rm{s}}$-band magnitudes are 
upper limits due to nearby source contamination. 
In the SIRIUS data, while the $K_{\rm s}$-band is unavailable due to 
saturation, the magnitudes of (\textit{J}, \textit{H})\,$=$\,(15.53, 10.05) 
were derived with photometric errors of $\sim$\,0.01~mag. 

\begin{figure}[!htb]
  \begin{center}
    \FigureFile(85mm,85mm){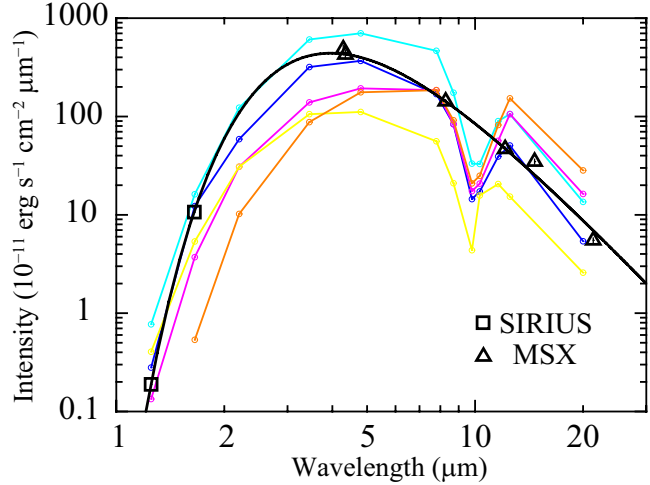}
  \end{center}
  \caption{SED constructed from SIRIUS (open squares) and MSX 
    (open triangles) photometry. 
  The solid curve shows the best-fit blackbody model with 
  $T_{\rm BB}\,=\,980$~K, $A_{\rm{V}}\,=\,$31~mag and 
  $L_{\rm{bol}}\,=\,8.3\times10^4$~\LO. 
  SEDs of the eponymous Quintuplet cluster members \citep{okuda90, figer99a} 
  are also shown with colors for comparison.}
  \label{fg:nirsed}
\end{figure}

In the MIR bands, the source is in the linearity range in the MSX photometry, 
but is too bright in Spitzer images, 
from which no meaningful photometry was obtained. 
Figure~\ref{fg:nirsed} shows the spectral energy distribution (SED) in the 
1.2--30~$\mu$m band. 
We fitted the data with a single-temperature blackbody and an 
assumed extinction proportional to the inverse square of the wavelength. 
We found a best-fit blackbody temperature ($T_{\rm{BB}}$) of 
980\,$\pm$\,20~K, with a visual extinction ($A_{\rm V}$) of 31\,$\pm$\,1~mag, 
and a bolometric luminosity 
($L_{\rm{bol}}$) of (8.3\,$\pm$\,0.5)$\times10^4~\LO
\left(d/8.0~\rm{[kpc]}\right)^2$. 

\section{Discussion}
\subsection{Location of Extinction Matter}
Adopting $N_{\rm H}/A_{\rm V}=1.79\times10^{21}$~cm$^{-2}$~mag$^{-1}$ 
\citep{predehl95}, $A_{\rm V}=31$~mag is converted to 
$N_{\rm H}\sim$\,5.6$\times10^{22}$~cm$^{-2}$. 
This is far smaller than that determined with the X-ray of 
$N_{\rm H}\sim$\,2.4$\times10^{23}$~cm$^{-2}$. 
Since the visual extinction of 31~mag is the typical value toward the Galactic 
center \citep{catchpole90}, the excess X-ray absorption is probably due 
to local obscuring matter that radiates the detected IR emission. 
If such an obscuring matter with solar abundance spherically surrounds 
the X-ray source, the 6.4~keV line with an EW of $\sim200$~eV should be 
detected \citep{inoue85}. 
The lack of 6.4~keV line (EW\,$<$\,50~eV) requires either that the iron 
abundance is $\lesssim0.25$~solar or that the local matter is concentrated 
in front of the source along the line of sight. 

\subsection{Nature of the Source}
\subsubsection{WR Binary Origin}
The SED of CXOGC\,J174645.3--281546 in the IR bands is very similar to 
those of the eponymous Quintuplet cluster members (figure~\ref{fg:nirsed}). 
The five stars have cool ($T\sim$\,1000~K) spectra and are bright 
($\sim$\,7~mag in the $K$-band). 
The deep absorption at $\sim\,$10~$\mu$m is due to interstellar silicate. 
The lack of either emission lines or intrinsic absorption 
features \citep{okuda89, figer99a} allows no spectral classification, 
though their luminosity of $\sim$\,10$^5$~$\LO$ 
corresponds to those of supergiant stars. 
These intriguing stars were recently spatially resolved \citep{tuthill06}.
Two (GCS\,3--2 and GCS\,4 of \cite{nagata90}, or Q2 and Q3 of \cite{moneti01}) 
out of five showed beautiful pinwheel nebulae of dust plume, which are 
seen in a few WC (carbon rich Wolf-Rayet) stars \citep{tuthill99, monnier99}. 
Circumstellar dust emission with temperatures of 700--1700~K is a 
common character of WC stars \citep{williams87}. 

Interestingly, the two sources that exhibit the pinwheel morphology 
are the brightest hard X-ray sources \citep{law04, wang06} among 
the Quintuplet members. 
Since both the pinwheel dust plumes and the hard X-ray emissions indicate 
fast wind-wind collision \citep{tuthill99, monnier99, oskinova03}, 
this coincidence is not probably accidental. 
The summed spectrum of three X-ray emitting sources 
(GCS\,3-2, GCS\,4, and source D of \cite{nagata90}) in the Quintuplet 
cluster is very hard ($kT\sim$\,9~keV) and shows a hint of iron K emission 
line, but the luminosity is smaller by $\sim$\,2 orders of magnitude  
(7.6$\times10^{32}$~erg~s$^{-1}$ in the 0.3--8.0~keV band; \cite{wang06}) 
than CXOGC\,J174645.3--281546. 
In the colliding wind binary scenario, the X-ray luminosity scales as 
$\lx\propto D^{-1}$ \citep{stevens92}, where $D$ is 
the binary orbital separation. 
The large X-ray luminosity of CXOGC\,J174645.3--281546 indicates the close 
separation, 
and the moderate flux variation having a factor of $\sim$\,2 can be 
interpreted as the variation of the separation due to the orbital motion. 

\begin{figure}[h]
  \begin{center}
    \FigureFile(85mm,85mm){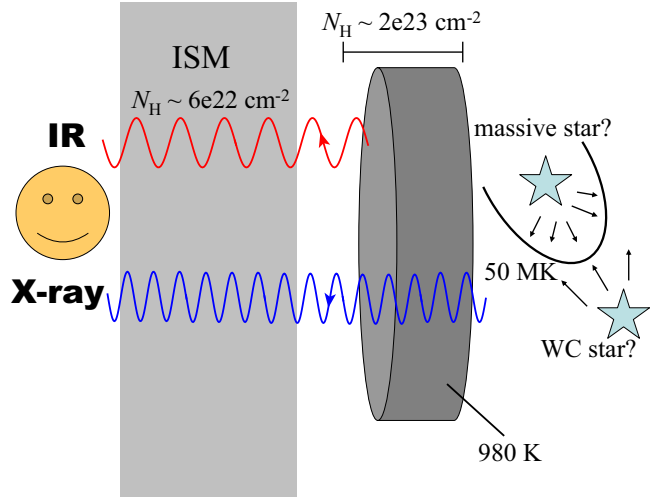}
  \end{center}
  \caption{Cartoon of the source system. Winds from the primary WC star and 
    the companion massive star collide to produce X-rays. 
    The obscuring material of $N_{\rm H}\sim$\,2$\times10^{23}$~cm$^{-2}$ 
    around the binary system is heated to $T\sim$\,980~K by 
    photospheric emission to emit IR. 
    Whereas the IR emission suffers only interstellar medium (ISM) extinction, 
    the X-ray emission is subject to additional intrinsic absorption. 
  } 
  \label{fg:cartoon} 
\end{figure}

Some of evolved massive~star+OB~star systems have very large X-ray luminosity 
up to $\sim$10$^{35}$~erg~s$^{-1}$ in the 0.5--10.0~keV band with thermal 
spectra with a temperature of $\gtrsim$\,2~keV and an iron abundance of 1--2 
($\eta$ Carinae: \cite{tsuboi97}, \cite{hamaguchi07}; WR\,140: 
\cite{koyama94}, \cite{pollock05}, 
A1\,S, A1\,N, and A2 in the Arches cluster; \cite{wang06}). 
The X-ray spectral property of CXOGC\,J174645.3--281546 coincides 
with those of these massive star binaries, though near- and mid-IR 
spectroscopy and high spatial resolution observation are essential for 
further constraint. 
Taking the discussion in \S4.1. into account, all of the observational results 
are consistent with the idea that CXOGC\,J174645.3--281546 is a 
WC star+massive star 
binary near the Galactic center, which has local obscuring 
matter (figure~\ref{fg:cartoon}). 

Because the initial mass of a WR star exceeds $\sim\,$40~\MO\ and the age is 
less than $\sim$6~Myr \citep{maeder87}, 
we naturally expect that WR stars are found in 
large young star clusters, assuming a usual initial mass function. 
Our source, however, seems to be isolated (figure~\ref{fg:irimage}). 
\citet{muno06a} and \citet{mauerhan07} also recently discovered 
such isolated massive stars in the Galactic center region. 
One possibility is that these stars were formed initially in a cluster, 
and the cluster dissipated due to the strong tidal force of the Galactic 
center \citep{portegies02a}. 
Future IR and X-ray observations with a higher spatial resolution and 
sensitivity may reveal that either this source is really a cluster member, 
or truly an isolated star. 

\subsubsection{Other Possible Origins}
The most popular class of hard X-ray sources that exhibits 
iron K-shell emission lines is CV. 
The X-ray luminosity of CXOGC\,J174645.3--281546 of 
$3\times10^{34}$~erg~s$^{-1}$ is at the brightest end of CVs. 
The bolometric luminosity of 
$\sim10^5~\LO\left(d/8.0~\rm{[kpc]}\right)^2$ is, however far larger 
than CVs, because most of the optical companions of CVs are late-type 
main-sequence stars. 
The spectra of CVs are characterized by hard continuum, with three 
iron K-shell emission lines at 6.4~keV, 6.7~keV, and 6.97~keV. 
The equivalent widths of both the 6.4~keV and 6.7~keV lines are, however,
around 100--200~eV, although in some cases that of the 6.7 keV line shows an 
exceptionally high value possibly due to resonance scattering \citep{terada01}.
The 6.97~keV/6.7~keV flux ratio of CVs is larger than 0.1, indicating that 
the ionization temperature is higher than $\sim\,$4~keV 
\citep{ezuka99, hellier04, rana06}. 
CXOGC\,J174645.3--281546 has a significantly larger equivalent width of 
the 6.7~keV line ($\sim\,1\,$keV) and lower flux ratio of 
the 6.97~keV/6.7~keV lines than those of CVs. 
The upper limit of 50~eV for the 6.4~keV line, on the other hand, 
can not give any constraint on the possibility of a CV. 
Altogether, CXOGC\,J174645.3--281546 is very unlikely to be a CV.

The X-ray spectra of YSOs are characterized by a thin thermal plasma
with a temperature of 1--5~keV and a metal abundance of around 0.3~solar
in the quiescent phase \citep{feigelson02, imanishi03, ozawa05}.
The CXOGC\,J174645.3--281546 iron abundance of 0.8\,$\pm$\,0.1~solar is higher
than the typical YSO value of 0.3~solar, but is not exceptional.
Although the ratio of the X-ray and bolometric luminosity 
($L_{\rm X}/L_{\rm bol}$) of $10^{-5}$ is in the range of low-mass YSOs 
\citep{imanishi01, feigelson02}, 
the absolute luminosities are too high for a YSO if CXOGC\,J174645.3--281546 
is located at 8~kpc. 
The possibility still remains that CXOGC\,J174645.3--281546 is a 
foreground YSO in a dense molecular cloud. 

\subsection{Contribution to the Galactic Center 6.7~keV Line}
The 6.7~keV line flux from CXOGC\,J174645.3--281546 is 
$2.5\times10^{-13}$~erg~s$^{-1}$~cm$^{-2}$. 
The total flux of iron K-lines in the Galactic center region is 
$\sim1.3\times10^{-10}$~erg~s$^{-1}$~cm$^{-2}$ \citep{yamauchi90}. 
Since this value includes the 6.4~keV and 6.97~keV lines, 
it overestimates the 6.7 keV flux by a factor of 2--3. 
The detected point sources account for $\sim10\%$ of the total flux of 
the 6.7~keV line \citep{wang02, muno04a}. 
Therefore, the 6.7~keV line of CXOGC\,J174645.3--281546 alone accounts for 
$\sim4$\% of that of the detected point sources and $\sim0.4$\% of that of 
the total diffuse flux. 
Such X-ray sources would comprise a substantial fraction of 6.7~keV 
line-emitting point sources in the Galactic center. 

\section{Summary}
\begin{enumerate}
\item We detected a strong 6.7~keV line with an equivalent width of 
$\sim$\,1~keV from CXOGC\,J174645.3--281546. 
The overall spectrum was very well fitted by a 
heavily absorbed (\nh$\sim$\,2.4$\times10^{23}$~cm$^{-2}$) 
3.8~keV thermal plasma with an iron abundance of $\sim$\,0.8~solar  
and a luminosity of $\sim$\, 3$\times10^{34}$~erg~s$^{-1}$ in the 
2.0--8.0keV band, assuming a distance of 8~kpc. 
\item We also analyzed the archived data of Chandra and XMM-Newton, and 
found that the X-ray flux spanning $\sim$\,6~years shows year-scale time 
variability having a factor of $\sim$\,2. 
\item The probable counterpart in the IR bands is very bright 
($L_{\rm bol}\sim$\,$10^{4.9}$~\LO~at 8~kpc), and has a cool 
($T_{\rm BB}\sim$\,1000~K) SED, which is similar to those of 
the eponymous members of the Quintuplet cluster. 
\item The X-ray spectral and temporal properties and the IR SED 
are fully consistent with a WC~star+massive~star binary system. 
\end{enumerate}
\bigskip

Y.\,H. and M.\,T. acknowledge financial support 
from the Japan Society for the Promotion of Science.
The work is financially supported by the grants-in-aid for a 
21st century center of excellence program 
``Center for Diversity and Universality in Physics'' and No. 18204015 
by the Ministry of Education, Culture, Sports, Science and Technology 
of Japan. 
This research has made use of data obtained from the Data ARchive and
Transmission System at ISAS/JAXA, the High Energy Astrophysics 
Science Archive Research Center at the NASA/Goddard Space Flight Center, 
and the Two Micron All Sky Survey, which is a joint project of 
the University of 
Massachusetts and the Infrared Processing and Analysis Center/California 
Institute of Technology, funded by the National Aeronautics and 
Space Administration and the National Science Foundation. 
MSX was funded by the Ballistic 
Missile Defense Organization with additional support from NASA 
Office of Space Science.  This research has also made use of the 
NASA/ IPAC Infrared Science Archive, which is operated by the 
Jet Propulsion Laboratory, California Institute of Technology, 
under contract with the National Aeronautics and Space 
Administration.

\end{document}